\documentclass[11pt]{article} 

\usepackage[utf8]{inputenc} 

\usepackage{geometry} 
\usepackage[pdftex]{graphicx} 
\usepackage{bmpsize}

\usepackage{booktabs} 
\usepackage{array} 
\usepackage{paralist} 
\usepackage{verbatim} 
\usepackage{subfig} 

\usepackage{natbib}

\usepackage[displaymath, mathlines]{lineno}

\usepackage{amsmath}
\usepackage{amssymb}
\usepackage{mathrsfs}
\usepackage{authblk}
\usepackage{fancyhdr} 
\usepackage{sectsty}
\allsectionsfont{\sffamily\mdseries\upshape} 

\usepackage[nottoc,notlof,notlot]{tocbibind} 
\usepackage[titles,subfigure]{tocloft} 

\title{Big-data approach in abundance estimation of non-identifiable animals with camera-traps\\ at the spots of attraction}
\author{Evgeny Ivanko}
\affil{Krasovskii Institute of Mathematics and Mechanics,\\ Ural Branch, Russian Academy of Sciences;\\ Ural Federal University\\ \ \\
\textit{S. Kovalevskoi 16, Ekaterinburg, 620990, Russia}\\
\textit{evgeny.ivanko@gmail.com}}

\date{}

\begin{document}


\maketitle

\begin{abstract}

Camera-traps is a relatively new but already popular instrument in the estimation of abundance of non-identifiable animals. Although camera-traps are convenient in application, there remain both theoretical complications such as spatial autocorrelation or false negative problem and practical difficulties, for example, laborious random sampling. In the article we propose an alternative way to bypass the mentioned problems. 

In the proposed approach, the raw video information collected from the camera-traps situated at the spots of natural attraction is turned into the frequency of visits, and the latter is transformed into the desired abundance estimate. The key for such a transformation is the application of the correction coefficients, computed for each particular observation environment using the Bayesian approach and the massive database (DB) of observations under various conditions.

The main result of the article is a new method of census based on video-data from camera-traps at the spots of natural attraction and information from a special community-driven database.

The proposed method is based on automated video-capturing at a moderate number of easy to reach spots, so in the long term many laborious census works may be conducted easier, cheaper and cause less disturbance for the wild life. Information post-processing is strictly formalized, which leaves little chance for subjective alterations. 
However, the method heavily relies on the volume and quality of the DB, which in its turn heavily relies on the efforts of the community.
There is realistic hope that the community of zoologists and environment specialists could create and maintain a DB similar to the proposed one. Such a rich DB of visits might benefit not only censuses, but also many behavioral studies.

\end{abstract}
\ \\


\section{Introduction}

Estimation of abundance (\citet{abun_gen2,abun_gen1}) of different species is an important part of duties of each nature reserve (NR). It helps to find the species in need of special attention and to balance the recreational load, to analyze the dynamics of populations and to reveal important ecological interconnections. Each NR can widen this list easily. 

The obvious way to estimate abundance is just to count all the individuals of the given species within the region of interest. However, the situation where such a direct method can be applied in practice is very rare. 


All other techniques may be considered indirect, where the researcher estimates the number of individuals under some limited conditions (both in space and time) and extrapolates the result. 

For the purposes of this article, it is convenient to divide the indirect techniques into two large groups:
1) with and 2) without individual identification. The backbone of the first group is the capture-recapture approach (\citet{cap_recap1,abun_gen2,abun_gen1}) broadened by modern identification methods such as
camera-trap estimation of individually identifiable animals (\citet{ident1,ident2,ident3,ident4,ident5,ident6,ident7,ident8,ident9,ident10})
or
microsatellite analysis (\citet{dna1,dna2,dna3,dna5,dna6,dna7,dna8}) of the collected biomaterial.
The capture-recapture method is very popular and effective, however, not many species have natural individual coloring or shape marks (like tiger's stripes or deer's antlers) allowing easy and practical application of camera-traps. For non-identifiable animals, the application of the capture-recapture method (both in the ``traps'' and ``camera-traps'' variations) is connected with considerable marking efforts and stress for the animals. DNA analysis remains a relatively difficult and expensive procedure, in addition, the process of samples collection is laborious and can hardly be facilitated. Summarizing, the techniques of this group perfectly fit the species with natural marks; otherwise, the researcher has to resort to relatively expensive and difficult artificial marking.



In the second group of estimation techniques corresponding to non-identifiable (or rather not easily identifiable) animals, distance sampling methods (\citet{abun_dist1,abun_dist2,abun_dist3}) dominate. Without individual identification, the researcher has only one type of data -- the frequency of the observations (either of the species itself or some kind of its traces). In this case, before solving the above-mentioned extrapolation problem, one needs to transform the frequency of the observations into an estimation of the density. 

Camera-traps is a relatively new but already popular instrument in the estimation of abundance of non-identifiable animals (\citet{noident5, noident2, noident1,noident4, noident3}). 
The method proposed in the article is a variation of distance sampling at point transects which are equipped with camera-traps. In the proposed approach highly attractive spots (below, we use the acronym AS for both singular and plural) like feeders or marking spots play the role of point transects, which allows to collect rich observation statistics and to cope with the false negative problem (\citet{false_neg2,false_neg1}). Another benefit of the AS usage is the reduced labor demands for the data collection. The corresponding overrated observation frequency is corrected by the coefficients computed basing on the Bayesian approach (see Section \ref{abundance}) applied to the specially prepared observation data (see Section \ref{db}).

For simplicity, below we consider abundance of solitary animals, however it is possible to proceed to social animals as soon as the average group size is known.






\section{AS: types, positioning and the corresponding area}
\label{posit}

All kinds of spots that attract the species of interest may be taken as AS, provided that three important conditions are met:

1) AS are placed inside the research area in such a way that almost every animal of the studied species living in (or regularly appearing in) the research area is aware of and periodically visits at least one AS. It is a separate difficult problem to place the AS correctly, but if we take, for example, feeders, NR encounters the positioning problem with the same criterion independently of camera-trap observations and solves it in some way thus we leave this discussion beyond the scope of the article.

2) There are no ``empty'' AS -- all the AS of the selected kind within the research area are equipped with the camera-traps so that any event ``an animal visits an AS'' is recorded. This condition is easy to fulfill if artificial AS such as feeders are used. Otherwise the experimenter should be confident that all the AS within the research area are discovered.


3) It is practically possible to use camera-traps to fix an event ``animal is at AS'' clearly and to measure the duration of this event. Technically, once a camera fixes an animal, it may turn off and recheck the animal presence periodically; thus we get a satisfactory estimate of the visit duration, while saving the battery. The event ``animal at AS'' may be defined differently depending on the type of AS and on the species under research, but, for a particular AS type and a species, this event should be clearly described  and commonly accepted by the professional community. For example, an elk may be considered staying at a feeder if and only if the distance between the elk and the feeder does not exceed  1 meter (or some other value accepted by the experts).








We mentioned the notion of ``research area'' above and even stipulated the restricting conditions on it. Let us clarify how this research area may be formally defined.

In our approach, camera-traps are used to measure the abundance over some area, after which the results are extrapolated to the desired part of the NR territory. The key to such an extrapolation is the density $D=N/S$, where $N$ is the estimated number of animals and $S$ is the ``area of the experiment.'' While $N$ is just a number resulting from observation by the methods discussed in the next section, the value of $S$ becomes ambiguous in case we consider animals in the wild. For example, we conduct observations in a forest without any fence and use just a few cameras (see Figure 1a--1d); which area should be preferred as a denominator in the expression for $D$? If there are more cameras (like in Figure 1e--1h), then the corresponding area appears a bit clearer. An enormous number of cameras (Figure 1i) is needed to define the corresponding area explicitly.





How can we rationally estimate the research area $S$? Let us suppose for a moment that we have an infinite net of more or less regularly positioned cameras on the surface beyond the finite set $C=\{\mathrm{AS}1,\ldots,\mathrm{AS}5\}$ of the real AS used in the experiment (see Figure  2a). In this case, the problem has a natural solution: we just need to surround the area ``corresponding to'' or ``proportional to'' the set $C$. Voronoi diagram (\citet{vor1,vordel}) give an elegant way to define the sought ``corresponding'' area formally (see Figure  2b).


In the observation experiment, we do not have any other AS except for the real, so new virtual neighbors should be constructed artificially. It is possible to propose many ways to expand  $C$ to an infinite net with some regularity; let us describe one of them briefly.

We start with the set $C$ of real AS, used in the observation experiment (see Figure  3a). At the first step, we build the Delaunay triangulation  (\citet{del1,vordel}) and calculate the average length $R$ of its edges (See Figure  3b). At the second step, we construct the $R$-equidistant of the set $C$ (see Figure  3c,d). The final third step consists of the following short algorithm (see Figure 3e): 1) start from an arbitrary point on the equidistant (a better choice is an intersection of two neighboring circles) $X1$; 2) construct the circle of radius $R$ from the center $X1$ and find $X2$ -- the intersection of this circle with the equidistant in the clockwise direction from the starting point; 3) repeat the process from the newly constructed point until we reach $X1$ from another side.

\section{Abundance estimation}
\label{abundance}

Suppose we have an unknown number $N$ of animals and $m$ stationary AS each of which is equipped with a camera-trap. The observation is conducted during the time $\mathbb{T}$ simultaneously for all the AS. 

The conditions of the observation may change during the experiment. The corresponding frequency and duration of visits as well as the resulted abundance estimate may vary significantly depending on the experiment condition. For example, a part of an observation experiment may pass under the rain, another part is conducted at night and so on. Let $W$ be the number of different conditions during the observation experiment. Each condition $\mathscr{A}^j$, where $j\in\overline{1,W}$,  is described by the sequence of states $\mathscr{A}^j=(\alpha_1^j,\ldots,\alpha_\Theta^j)$, lasts for the time $T(\mathscr{A}^j)\in[0,\mathbb{T}]$ and results in the estimated number of animals $N_j$. 

For example, if an experiment is conducted during both day and night and faces both dry and rainy weather in each time of day, then there are $W=4$ different conditions: 

$\mathscr{A}^1=(rain, day)$, 
$\mathscr{A}^2=(rain, night)$, 

$\mathscr{A}^3=(dry, day)$, 
$\mathscr{A}^4=(dry, night)$.

The weighted average among the estimations under all the encountered conditions is used for the following final expression:

\begin{equation}
\tag{eqn 1}
N\approx\sum_{j=1}^W\frac{T(\mathscr{A}^j)}{\mathbb{T}}N_j.
\end{equation}

Each $N_j$ is approximated as the total ``successful'' (animal at AS) observation time $T^*(\mathscr{A}^j)$ (summed over all the animals and all the cameras) divided by the ``unit'' time $T^0(\mathscr{A}^j)$ one average animal of the species under research spends at AS during the conditions $\mathscr{A}^j$

\begin{linenomath*}
\begin{equation}
\label{nj1}
\tag{eqn 2}
N_j\approx T^*(\mathscr{A}^j)/T^0(\mathscr{A}^j).
\end{equation}
\end{linenomath*}

The value of $T^*(\mathscr{A}^j)$ may be measured from the observation experiment 

\begin{linenomath*}
\begin{equation}
\tag{eqn 3}
T^*(\mathscr{A}^j)=\sum_{i=1}^m T_i^*(\mathscr{A}^j),
\end{equation}
\end{linenomath*}

where $T_i^*(\mathscr{A}^j)$ is the total ``successful'' observation time from the $i$-th camera. It must be noted that $T^*_i(\mathscr{A}^j)$ is summed up over all the animals independently, so if one elk spent 14 minutes at the $i$-th AS, then another one came and there were two elks during 7 minutes and then again only one remained for another 4 minutes, then $T^*_i(\mathscr{A}^j)=14+7+7+4$ minutes. 

For practical purposes, it is more convenient to use relative time in
 \eqref{nj1},

\begin{linenomath*}
\begin{equation*}
\label{chisl_na_camere}
\tag{eqn 4}
N_j\approx K(\mathscr{A}^j)/K^0(\mathscr{A}^j),
\end{equation*}
\end{linenomath*}
where 
$K(\mathscr{A}^j)=T^*(\mathscr{A}^j)/T(\mathscr{A}^j)$ and $
K^0(\mathscr{A}^j)=T^0(\mathscr{A}^j)/T(\mathscr{A}^j)$.

While the relative ``successful'' time $K(\mathscr{A}^j)$ can be measured directly from the experiment, the ``unit'' relative ``successful'' time $K^0(\mathscr{A}^j)$ should be approximated a priori. Note that if we rely on the Ergodic Hypothesis, $K^0(\mathscr{A}^j)$ may be regarded as the probability of one average animal to be seen at one average AS at a randomly selected moment $t_0\in[0,T(\mathscr{A}^j)]$ during the conditions $\mathscr{A}^j$. 

Rewriting this probability using the Bayes naive classifier (\citet{bayes}) with the assumption that for each fixed $j$ the states $\alpha^{j}_k, k\in\overline{1,\Theta}$ are pairwise independent, we get

\begin{linenomath*}
\begin{equation*}
\label{one_k0}
\tag{eqn 5}
K^0(\mathscr{A}^j)=K^0\frac{K^0(\alpha_1^{j})\ldots K^0(\alpha_\Theta^{j})}{F(\alpha_1^{j})\ldots F(\alpha_\Theta^{j})},
\end{equation*}
\end{linenomath*}
where 

\begin{itemize}
\item $K^0$ is the probability to see an animal of the given species independently from the observation conditions; in practice it is the total time one animal of the given species spends at AS in any conditions divided by the total time of the observation experiments where we intended to see that species; 
\item $K^0(\alpha_k^{j})$ is the probability to see an animal of the given species at AS during the state $\alpha_k^{j}$; in the context of computation, it is the total time of ``animal at AS'' during the state $\alpha_k^{j}$ divided by the total time devoted to the observation of the corresponding species during this state;
\item $F(\alpha_k^{j})$ is the frequency of the state $\alpha_k^{j}$, which may be expressed as the total time of observation during the state $\alpha_k^{j}$ divided by the overall time of observation.
\end{itemize}

The resulting formula for the abundance estimate may be written as follows

\begin{equation}
\tag{eqn 6}
\label{main}
N\approx\sum_{j=1}^{W} \frac{T(\mathscr{A}^j)}{\mathbb{T}}\frac{K(\mathscr{A}^j) F(\alpha_1^{j})\ldots F(\alpha_\Theta^{j})}{K^0 K^0(\alpha_1^{j})\ldots K^0(\alpha_\Theta^{j})}.
\end{equation}

The values $K^0$, $K^0(\alpha_k^{j})$ and $F(\alpha_k^{j})$ may be estimated by means of a special database (see Section \ref{db}) aggregating large amounts of information about the relative time one animal of the selected species spends at AS in different conditions. Now, let us proceed to the discussion of how this time may be computed. 

\section{Presence time estimation}
\label{presence}
Although the estimation of the time one animal spends at AS is definitely not an easy problem, it seems more tractable than the abundance estimation problem itself. The main benefit of the proposed Bayes approach \eqref{main} is that the more heterogeneous data we aggregate in the DB, the more exact abundance estimate we get in a wide variety of possible conditions. In this section we show how one can obtain the experimental information to improve the DB with the results of his local observations.

Within this section we suppose that the experiment conditions and the duration of observation are the same for all the AS, so $\mathscr{A}^j=\mathscr{A}$ and $T(\mathscr{A}^j)=T(\mathscr{A})=\mathbb{T}$.

Several techniques may be proposed for the measurement of the time one average unidentifiable animal spends at AS. The first and the most obvious is to identify the visitor, for example, using a GPS-collar (\citet{gps}), so we know exactly where the animal is. If there are $m'$ cameras and $N'$ GPS-collars, then, for the $k$-th animal with a collar, we have the following approximation for the ``unit'' relative ``successful time'':


\begin{linenomath*}
\begin{equation}
\tag{eqn 7}
\widetilde{K}^0(k,\mathscr{A})\triangleq\sum_{i=1}^{m'} \frac{\widetilde{T}_i^*(k,\mathscr{A})}{T(\mathscr{A})},
\end{equation}
\end{linenomath*}
where $\widetilde{T}_i^*(k,\mathscr{A})$ is the time that the $k$-th animal spent at the $i$-th AS during the conditions $\mathscr{A}$.

The resulting $K^0(\mathscr{A})$ is expressed as the average of the relative ``successful'' time among all the tracked animals

\begin{equation*}
\tag{eqn 8}
\label{k02}
K^0(\mathscr{A})=\frac{1}{N'}\sum_{k=1}^{N'}\widetilde{K}^0(k,\mathscr{A}).
\end{equation*}

The same technique works if we mark several animals in some way (see e.g. \citet{mark}) to be able to recognize them and to measure $\widetilde{T}^*_i(k,\mathscr{A})$ for each of them directly at the AS by camera-traps.

The third -- a bit more complicated -- approach is not connected with individual identification. Instead, the number of animals $N'$ which visit the involved $m'$ attractive spots is supposed to be known: either we take a semi-wild closed group with a priori known number of individuals or there exist some trusted methods to estimate the number of distinct animals visiting the AS (e.g. microsatellite analysis). We use the same logic as in \eqref{nj1} and \eqref{chisl_na_camere}, remembering that $\mathscr{A}^j=\mathscr{A}$ and $T(\mathscr{A}^j)= T(\mathscr{A})= \mathbb{T}$:

\begin{linenomath*}
\begin{equation}
\tag{eqn 9}
N'=\sum_{i=1}^{m'}\frac{T_i^*(\mathscr{A})}{T^0(\mathscr{A})}=\sum_{i=1}^{m'}\frac{T^*_i(\mathscr{A})/\mathbb{T}}{K^0(\mathscr{A})},
\end{equation}
\end{linenomath*}

whence

\begin{equation*}
\label{k03}
\tag{eqn 10}
K^0(\mathscr{A})=\frac{T^*(\mathscr{A})}{N'\mathbb{T}}.
\end{equation*}

\section{Database}
\label{db}
The DB (see Figure 4) contains five types of columns: 
\begin{enumerate}
\item species \textit{id} (e.g. Vulpes); 
\item mandatory experiment properties $\mathrm{A}=(\alpha_1,\ldots,\alpha_\Psi)$ (e.g. day time, season, weather, climatic zone  and so on); the results of the presence time estimation experiment (see Section \ref{presence}) may be added to the DB only if all the mandatory experiment properties are registered and the corresponding columns of the DB are filled;
\item optional experiment properties $\mathrm{B}=(\alpha_{\Psi+1},\ldots,\alpha_\Omega)$ (e.g. conditions of specific food reserves, distance from specific ecological zones and so on); 
\item duration $T(\mathscr{A})$ of observation under the particular conditions $\mathscr{A}=(\alpha_1^j,\ldots,\alpha_\Theta^j)$ (where $\mathscr{A}$ covers all the properties from $A$ and maybe some from $B$);

\item the corresponding average relative time $K^0(\mathscr{A})$ an individual animal spends at AS obtained from the presence time estimation experiment.
\end{enumerate}



There are three main reasons to interact with the proposed DB: read, write and enhance. Let us consider each of them briefly.

\subsection{Read}

The proposed DB gives all the necessary information for applying \eqref{main} in an   abundance estimation experiment conducted for the species $Sp$:
\begin{itemize}
\item $K^0$ is approximated by the average value of the column ``$K^0(\mathscr{A})$'' among all the rows with the column ``species'' equal to $Sp$;
\item $K^0(\alpha^{j}_k)$ is approximated by  the average value of ``$K^0(\mathscr{A})$'' among all the rows with ``species'' equal to $Sp$ and ``$\alpha_k$'' column equal to $\alpha^{j}_k$;
\item $F(\alpha^{j}_k)$ is approximated by  the summary observation time ``$T(\mathscr{A})$'' over all the rows with ``$\alpha_k$'' equal to $\alpha^{j}_k$ divided by the total observation time ``$T(\mathscr{A})$'' summed over the whole DB.
\end{itemize}


\subsection{Write}
Write operation is the simplest one to explain: the researcher conducts one of the presence time estimation experiments described in Section \ref{presence} and appends one new line to the DB, filling the columns according to the experiment's conditions and results.

As it was mentioned before, the properties $\mathrm{A}=(\alpha_1,\ldots,\alpha_\Psi)$ must be measured and filled, while the properties $\mathrm{B}=(\alpha_{\Psi+1},\ldots,\alpha_\Omega)$ are optional.

\subsection{Enhance}

There are two main roles of the DB users: researcher and expert. A researcher may propose: 1) new optional properties; 2) candidate optional properties to become mandatory; 3) to degrade some mandatory properties to optional; 4) to remove optional properties from the DB. A researcher may also vote for all the propositions made by the other researchers.

An expert is a researcher who decides the destiny of the properties according to the public opinion and statistics. If the variation of some property (with all the other properties averaged) does not result in a significant variation of $K^0(\mathscr{A})$, then this property is a candidate for degrading from the mandatory or removing from the optional properties. And vice versa, if $K^0(\mathscr{A})$ varies significantly depending on a particular optional property, then this property becomes a candidate for mandatory.

There is a lot of statistical methods to evaluate the significance of dependency and the decision threshold (\citet{stat1,stat2,stat3}), we will not discuss this topic here.







\section{Conclusion}

The proposed method is based on automated video-capturing at a moderate number of easy to reach spots, so in the long term many laborious census works may be conducted easier, cheaper and cause less disturbance for the wild life. Information post-processing is strictly formalized, which leaves little chance for subjective alterations. 

However, the method heavily relies on the volume and quality of the DB, which in its turn heavily relies on the efforts of the community. Public non-commercial solutions of complex problems (crowdsourcing) raises its popularity in such areas as software production (Freeware), investment (crowdfunding), scientific research (crowdsolving, civilian science). There is realistic hope that the community of zoologists and environment specialists could create and maintain a DB similar to the proposed one. Such a rich DB of visits might benefit not only censuses, but also many behavioral studies.


The author is open for collaboration, both aimed at the practical adaptation of the proposed method for specific regions and at the development of the census methods theoretically.

\section*{Acknowledgements}

This research was supported by Russian Foundation for Basic Research (grants 16-01-00649, 17-08-00682). The author wants to thank Dr. Nikolay Markov (Institute of Plants and Animals, Ural Branch of Russian Academy of Sciences) for the fruitful discussions and critics and Evgeny Larin (Nature Park ``Kondinskie Lakes'') for the practical insights.

\bibliographystyle{mee} 
\bibliography{mee_refs}

\newpage

\begin{center}
\begin{tabular}{c}
\begin{tabular}{cccc}
a\includegraphics[width=0.1\textwidth]{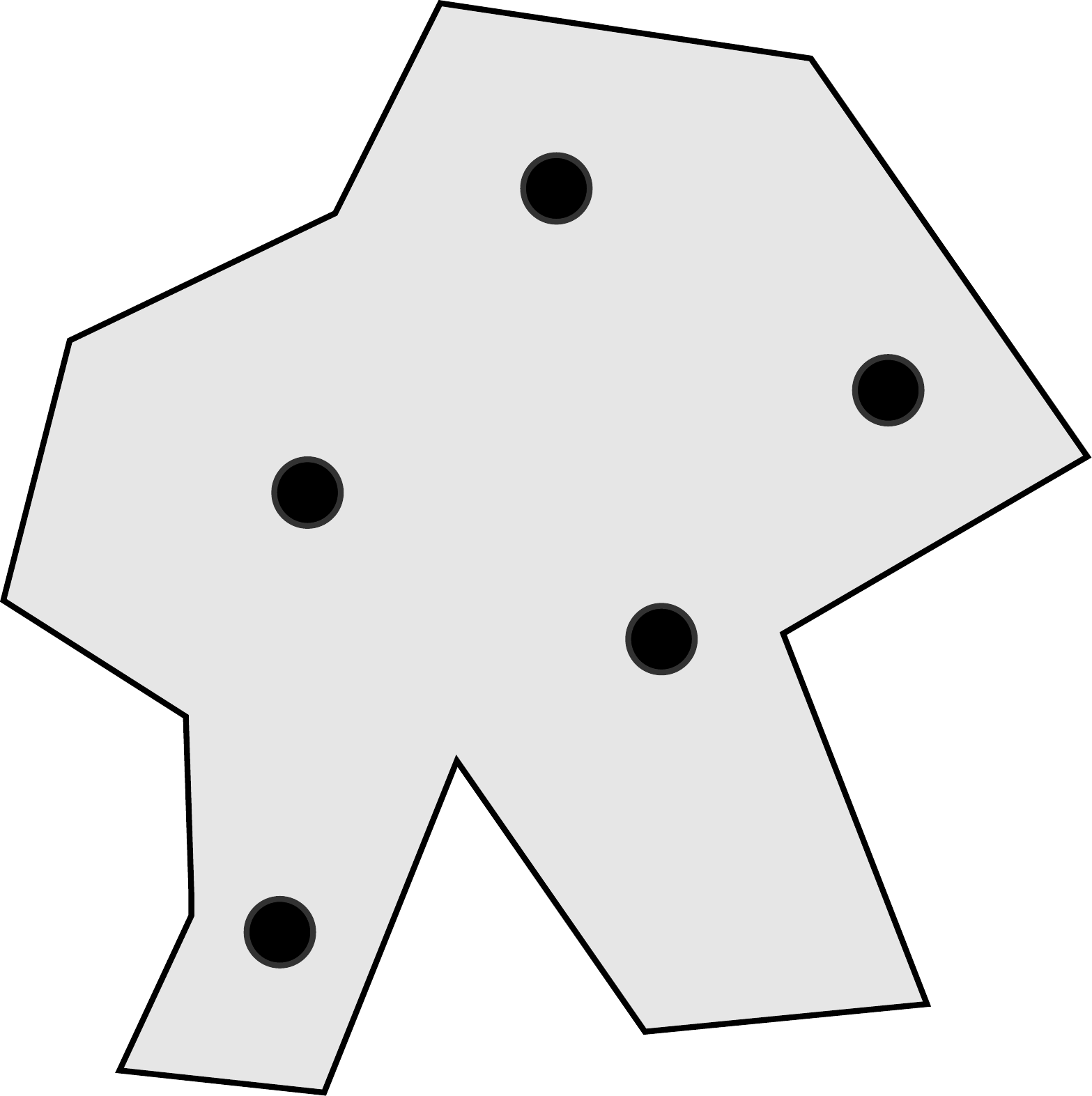}&
b \includegraphics[width=0.1\textwidth]{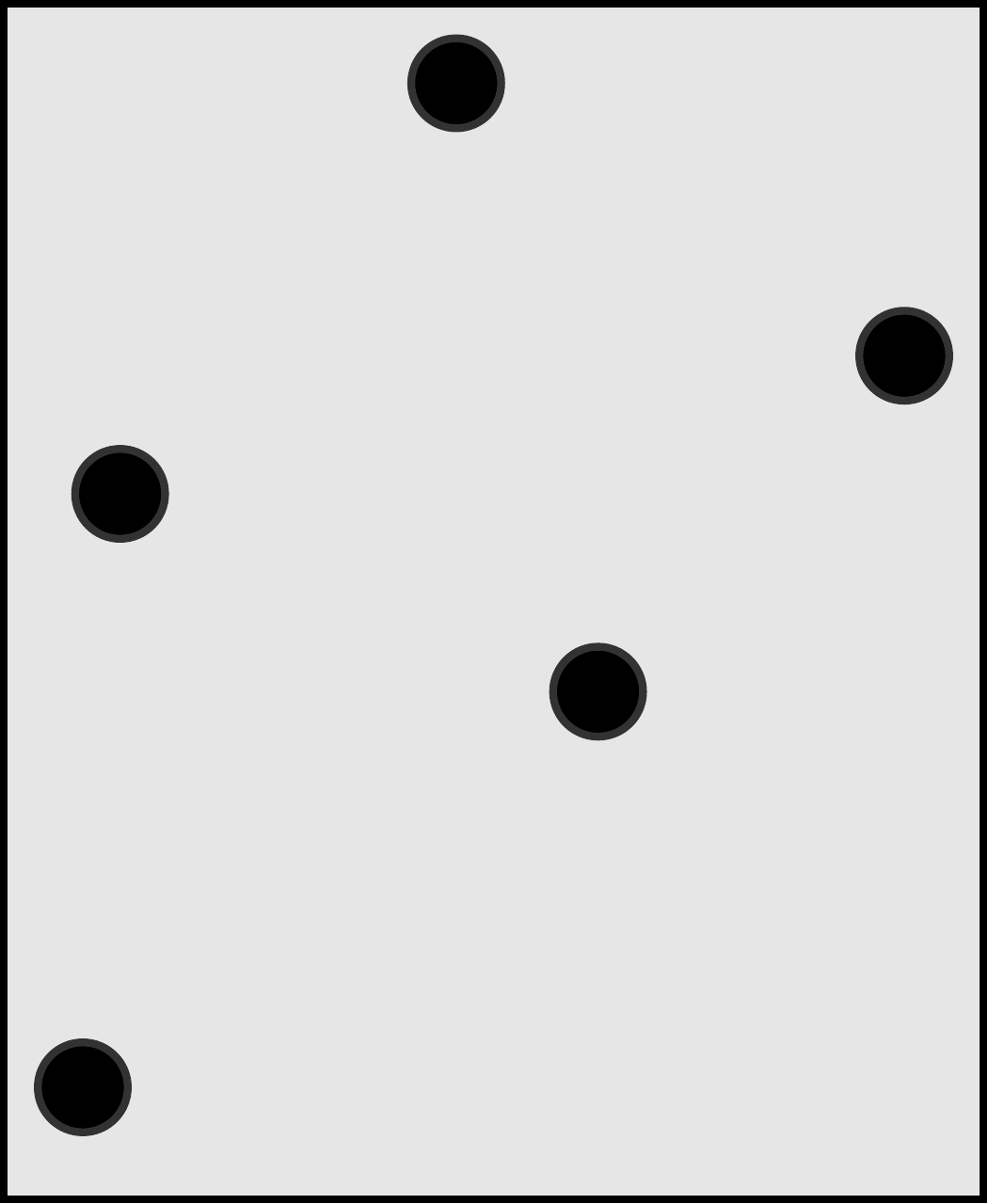}&
c \includegraphics[width=0.1\textwidth]{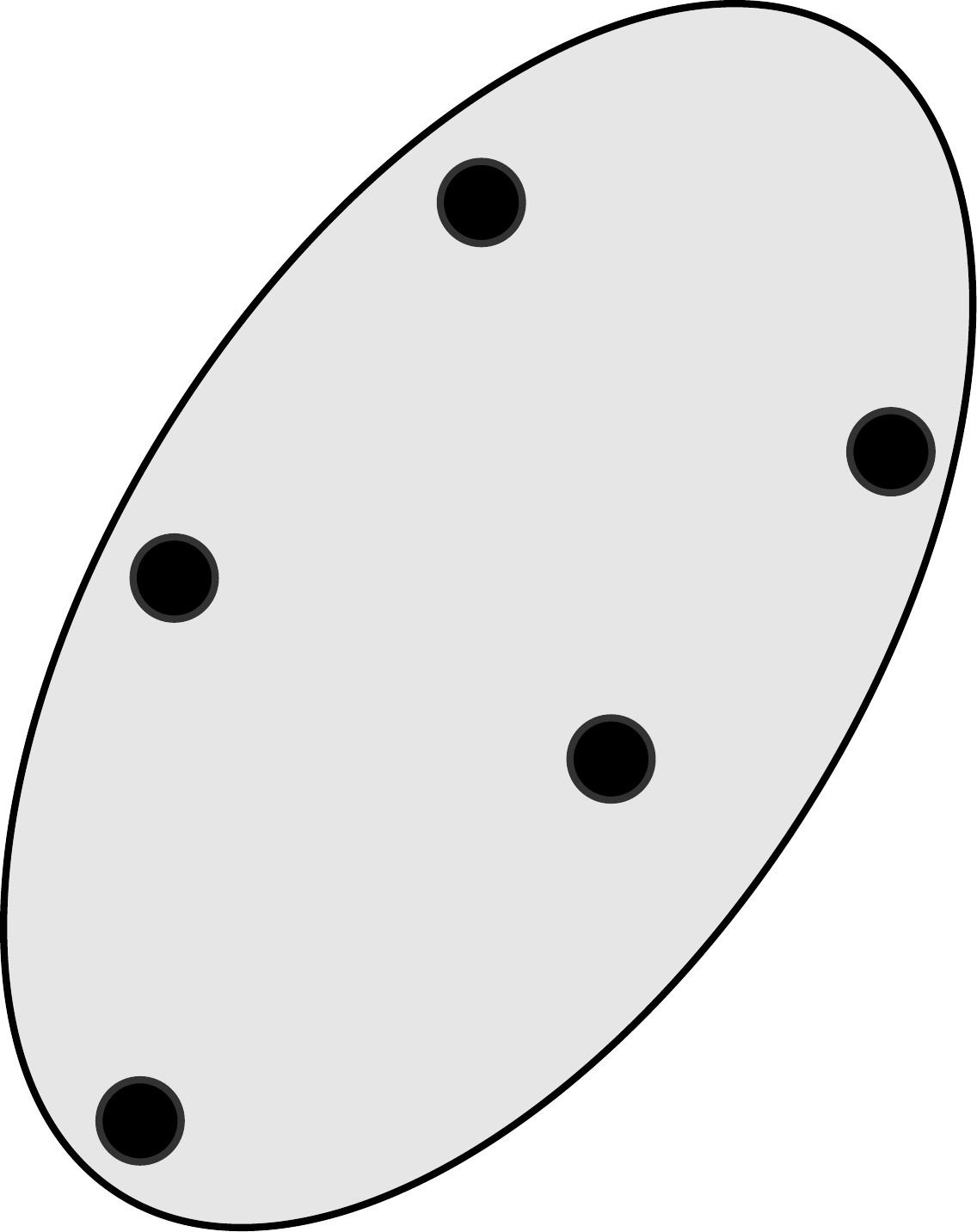}&
d \includegraphics[width=0.1\textwidth]{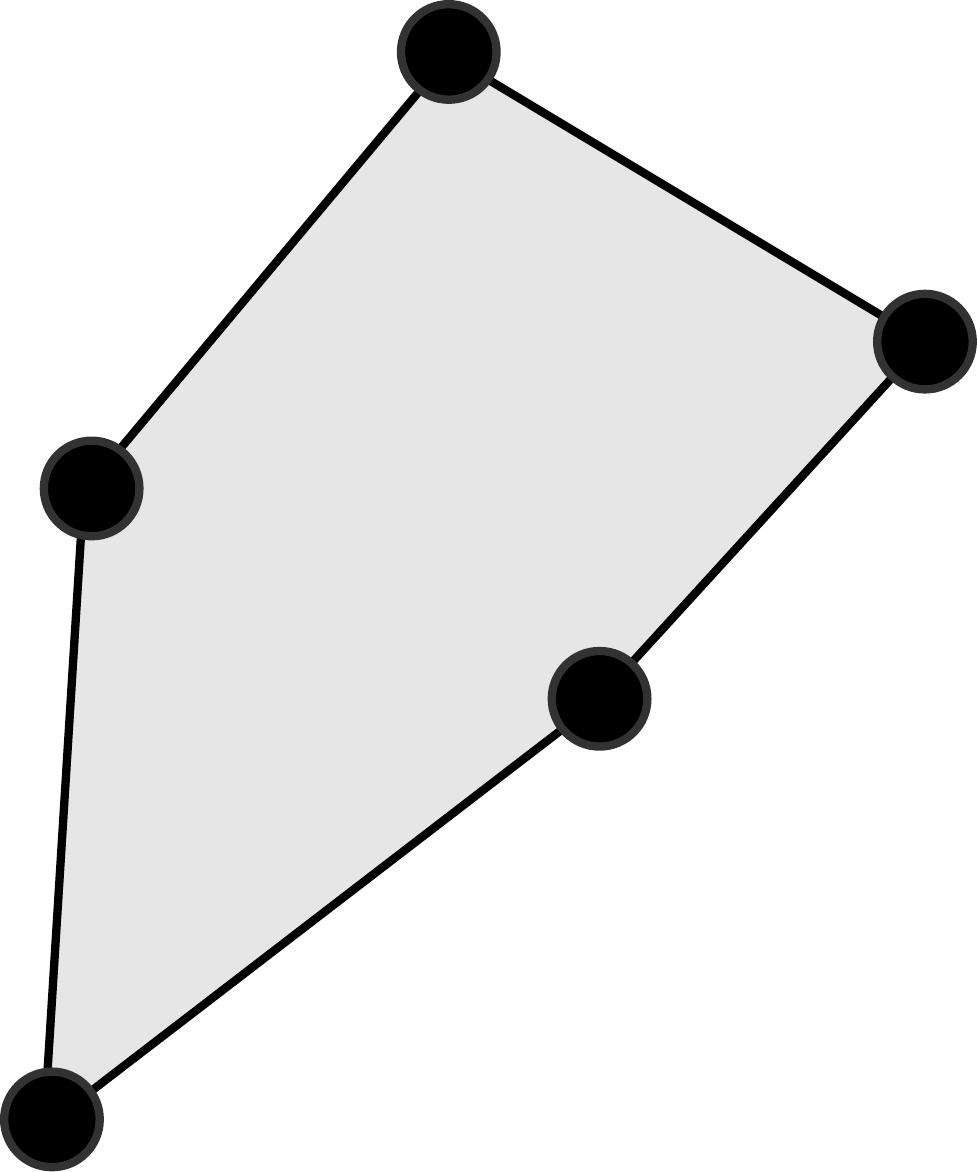}\\
e\includegraphics[width=0.1\textwidth]{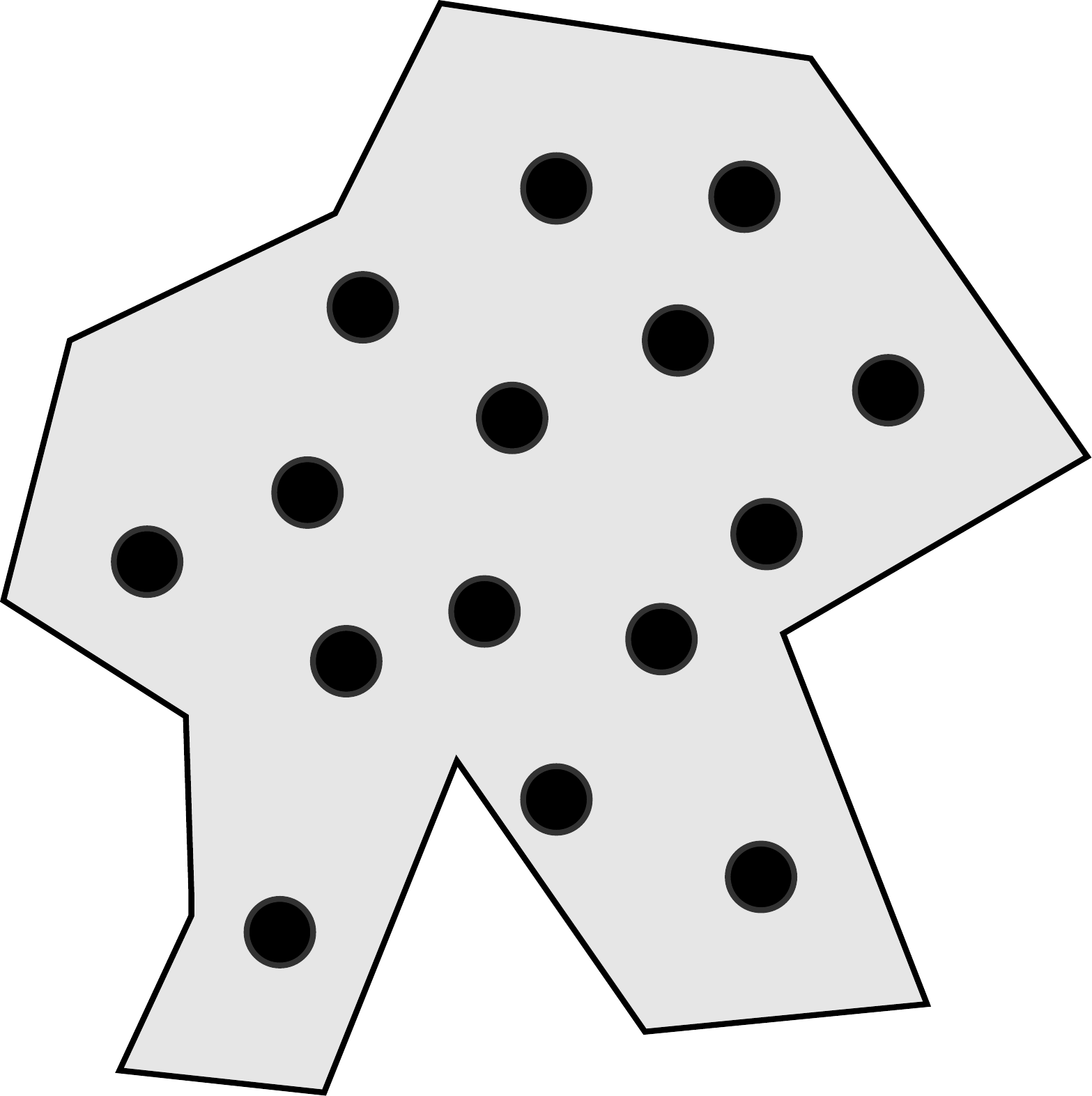}&
f \includegraphics[width=0.1\textwidth]{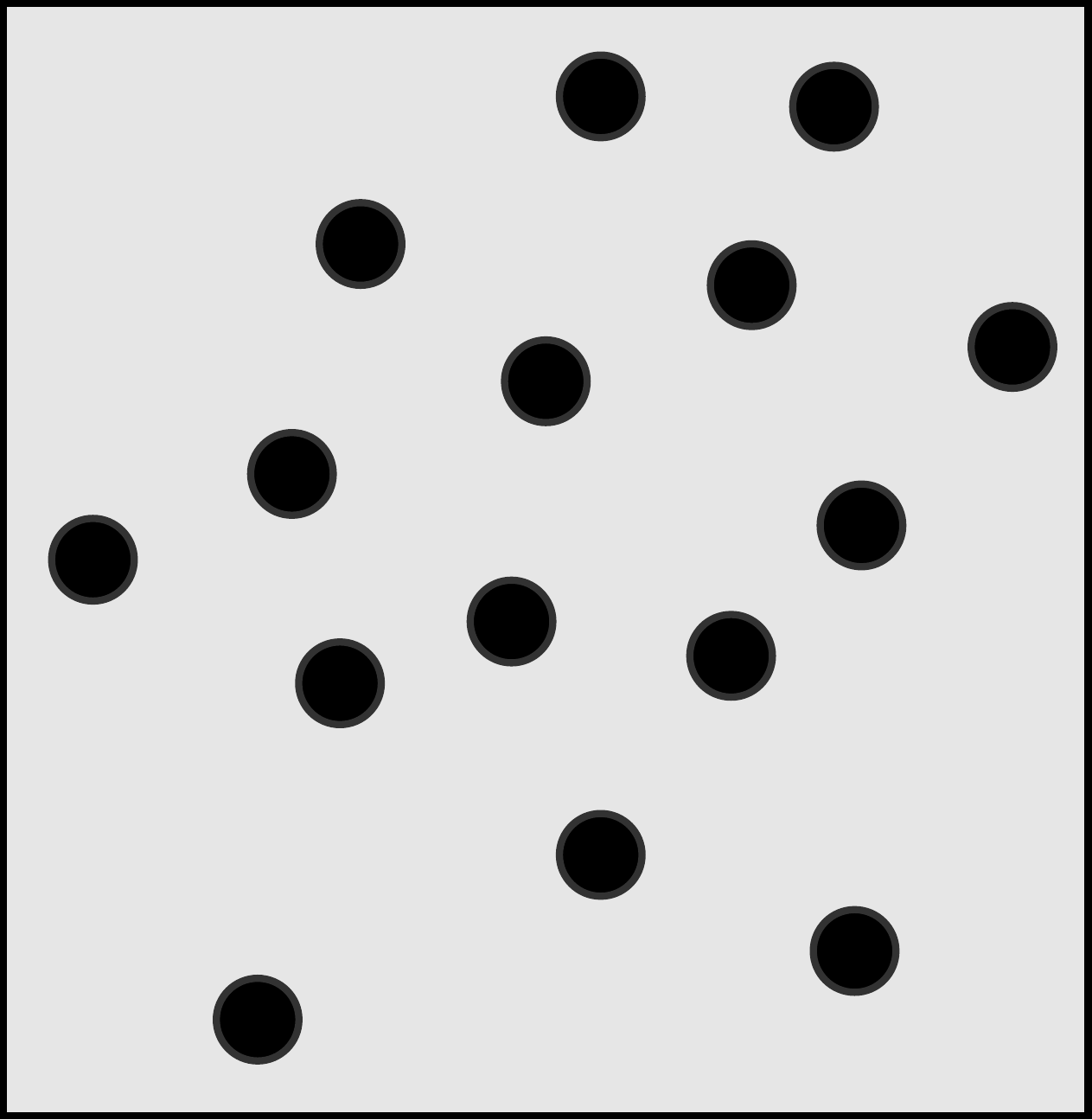}&
g\includegraphics[width=0.1\textwidth]{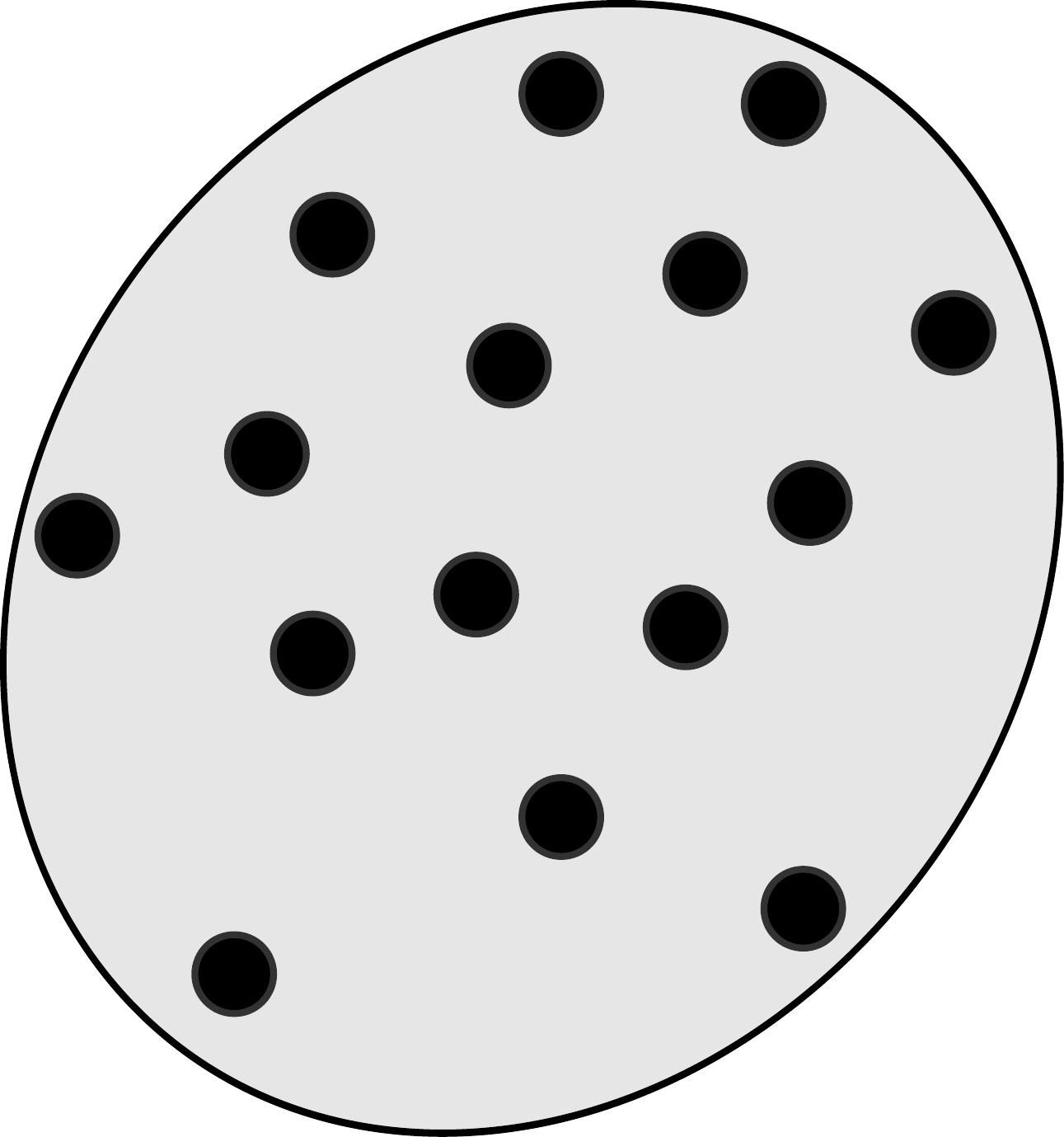}&
h\includegraphics[width=0.1\textwidth]{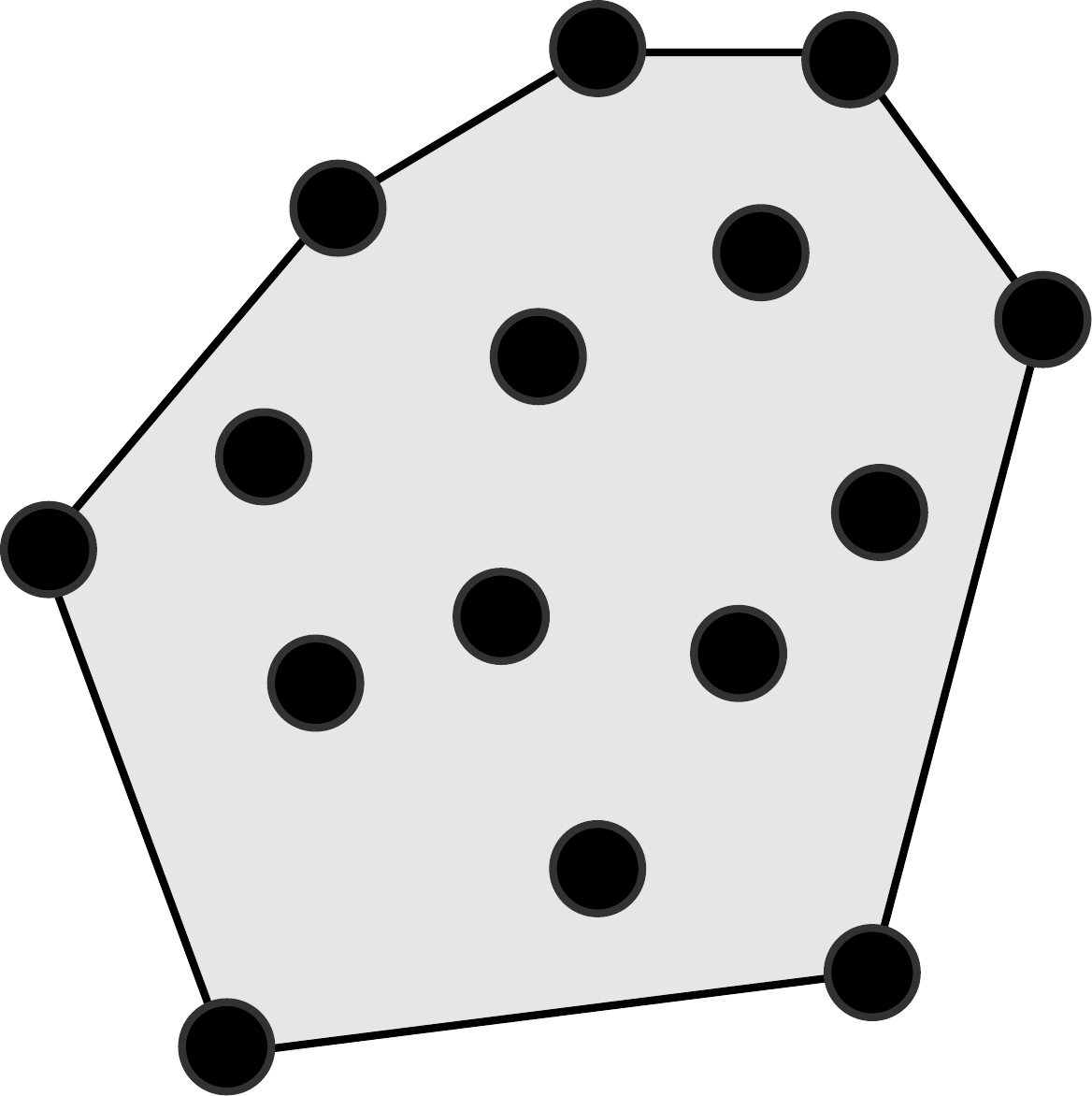}
\end{tabular}\\
\begin{tabular}{c}
i\includegraphics[width=0.2\textwidth]{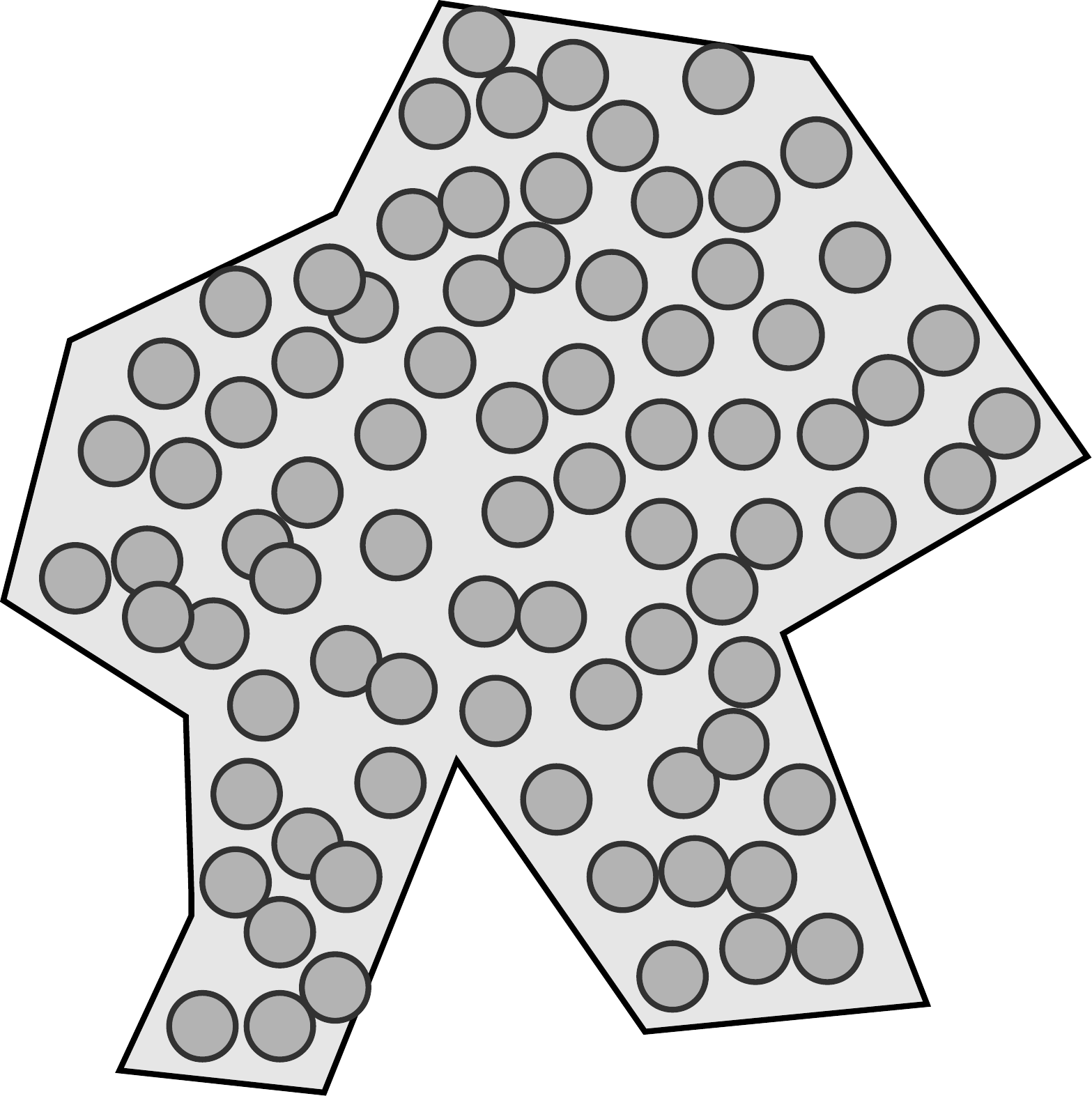}
\end{tabular}
\end{tabular}
\end{center}
{\bf Figure 1.} Which area better corresponds to the AS positions?
\vspace*{15px}

\newpage

\begin{center}
\begin{tabular}{c c}
a&b\\
\includegraphics[width=0.4\textwidth]{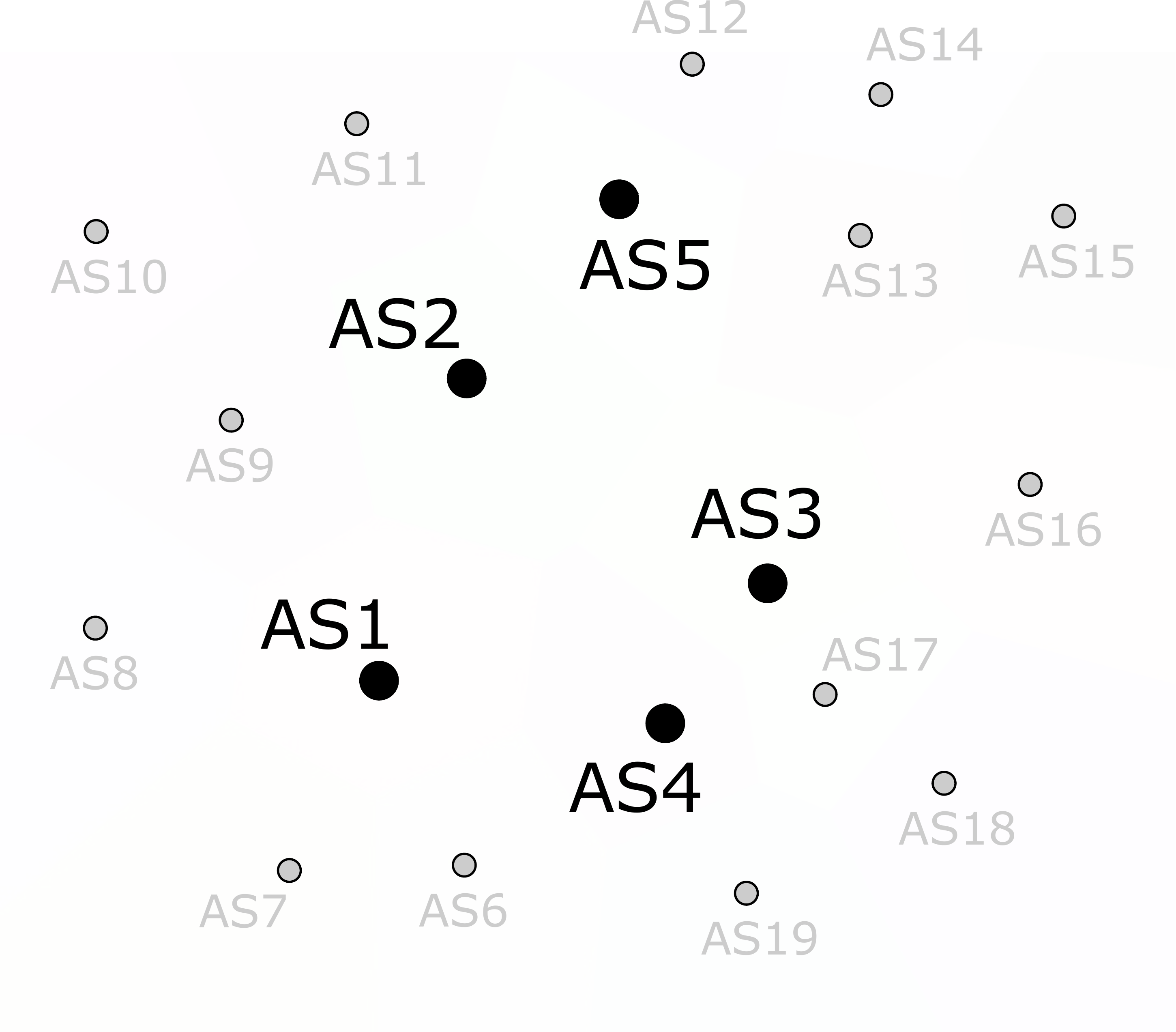}&
\includegraphics[width=0.4\textwidth]{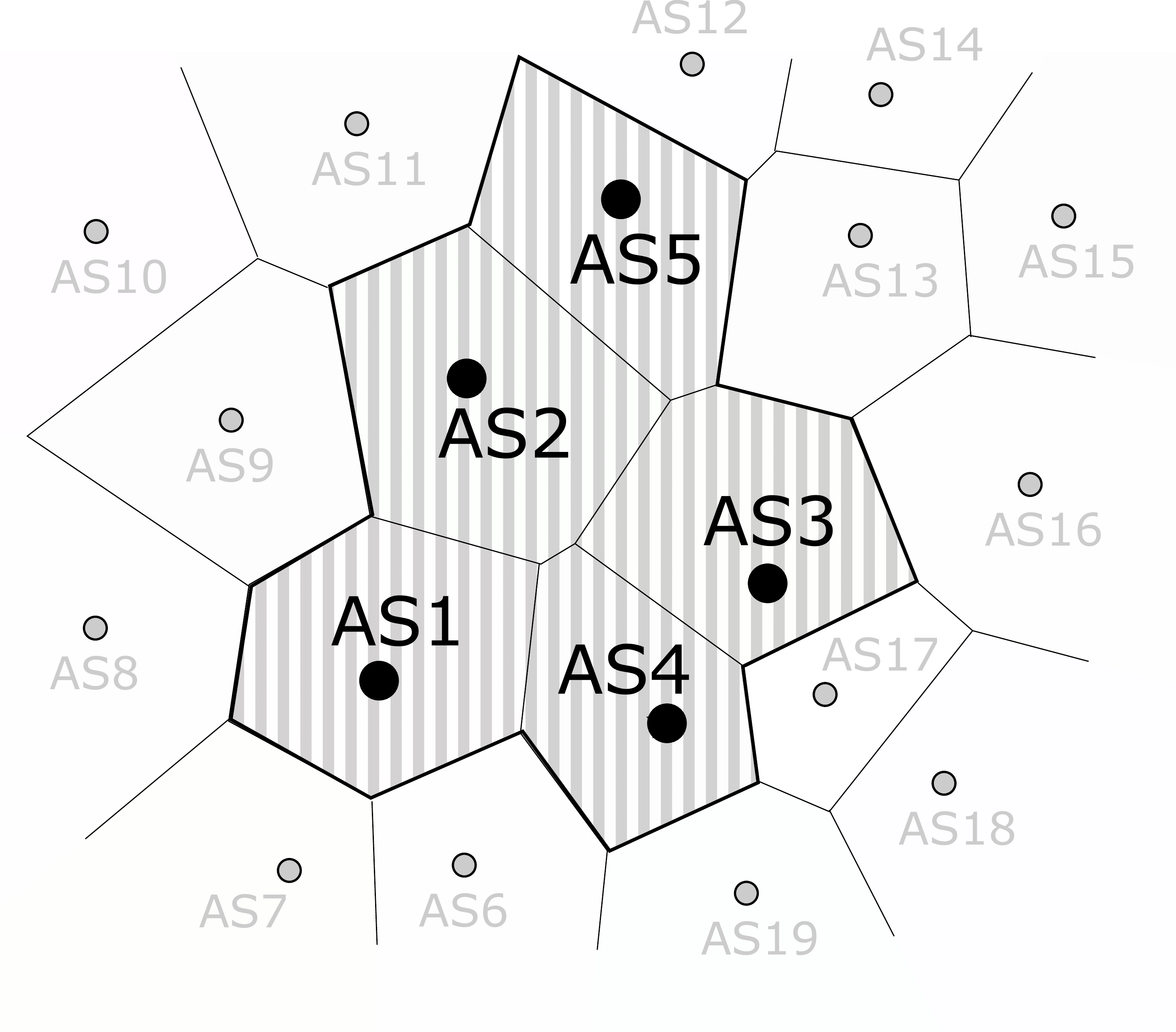}
\end{tabular}
\end{center}
\noindent{\bf Figure 2.} a) a part of an infinite AS net around AS1--AS5; b) the area, corresponding to the set AS1--AS5 according to the Voronoi diagram
\vspace*{15px}

\newpage
\begin{center}
\begin{tabular}{ccc}
a&b&c\\
\includegraphics[width=0.275\textwidth]{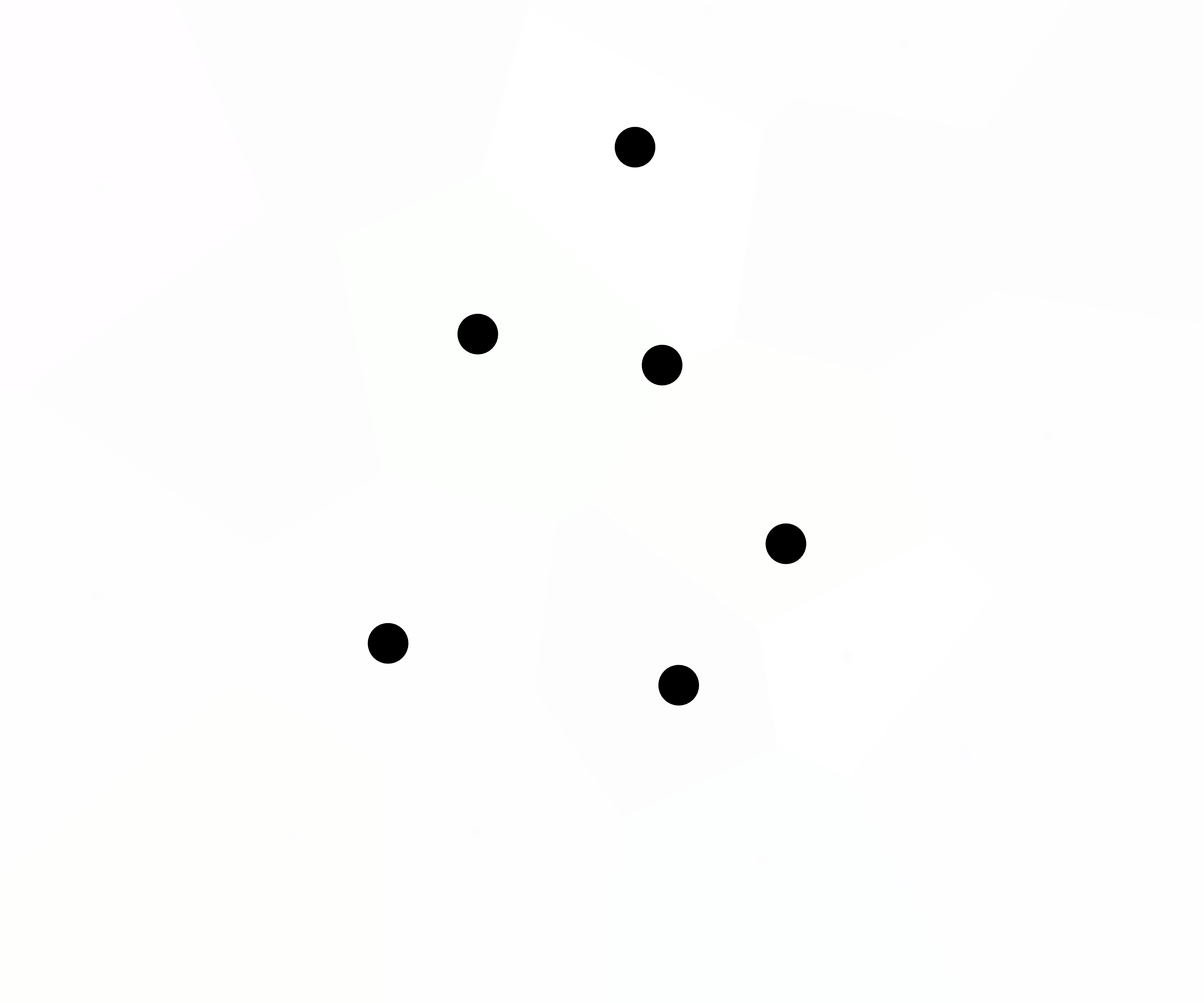}&
\includegraphics[width=0.275\textwidth]{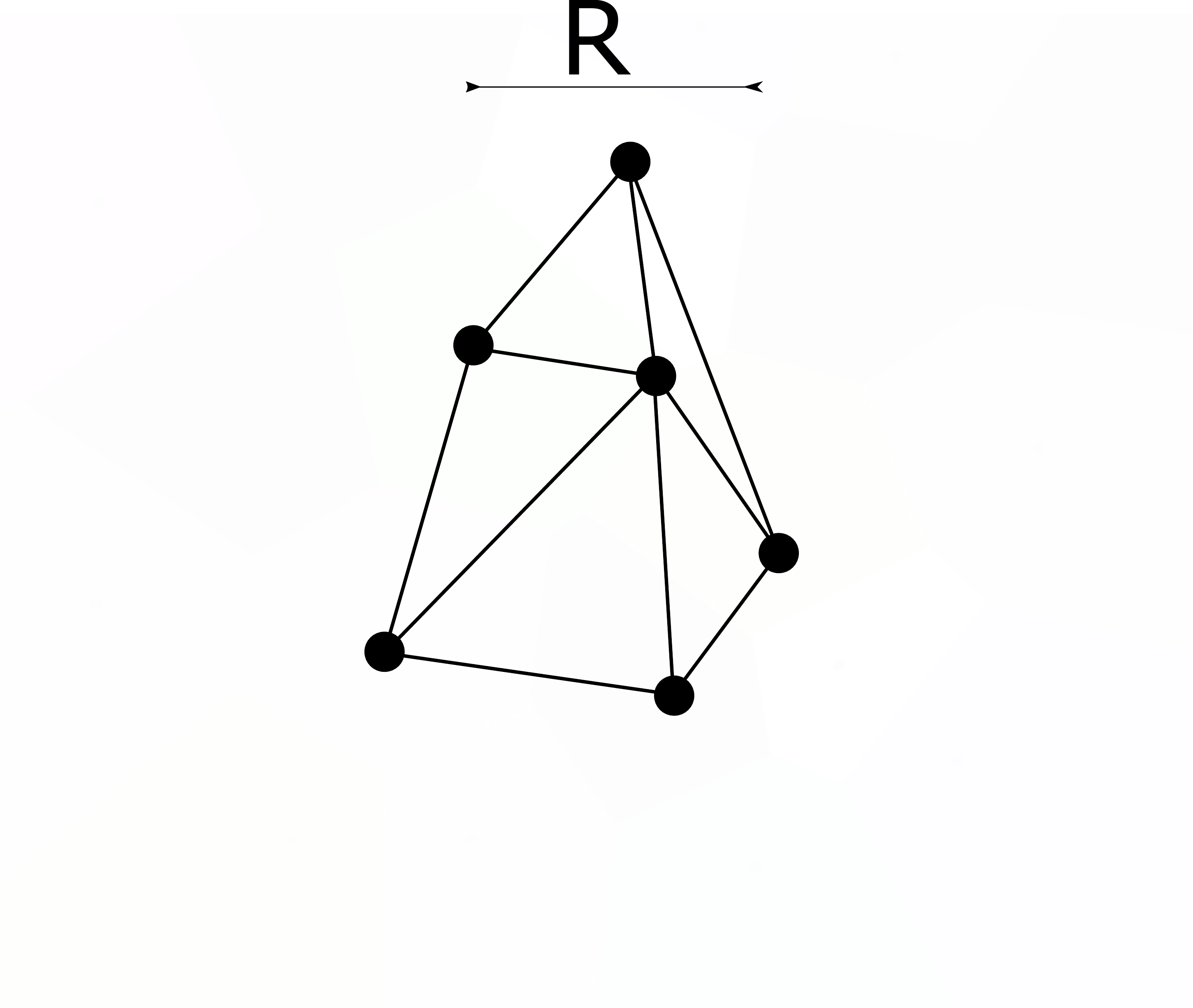}&
\includegraphics[width=0.275\textwidth]{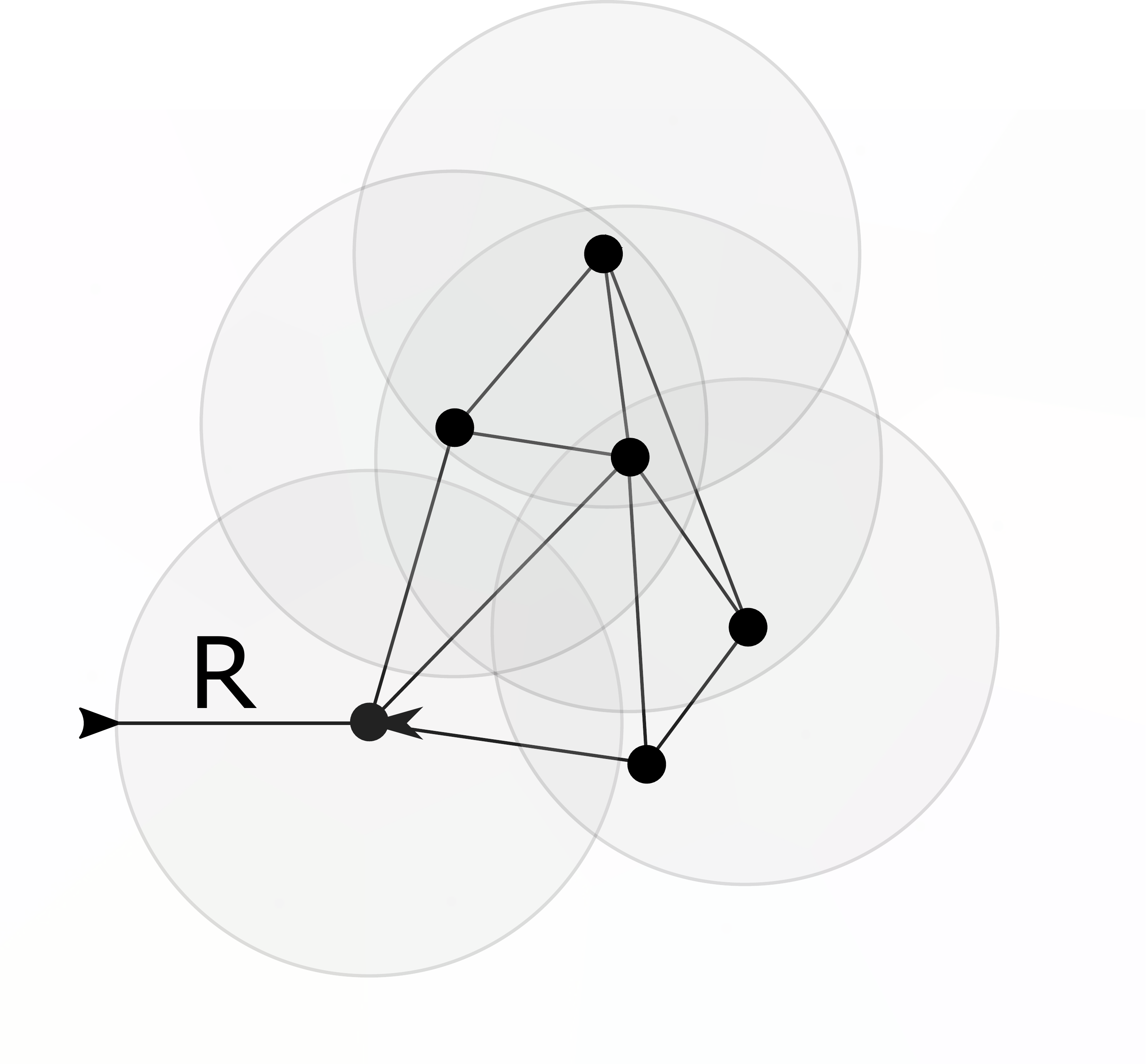}
\end{tabular}\\

\centering
\begin{tabular}{cc}
\includegraphics[width=0.275\textwidth]{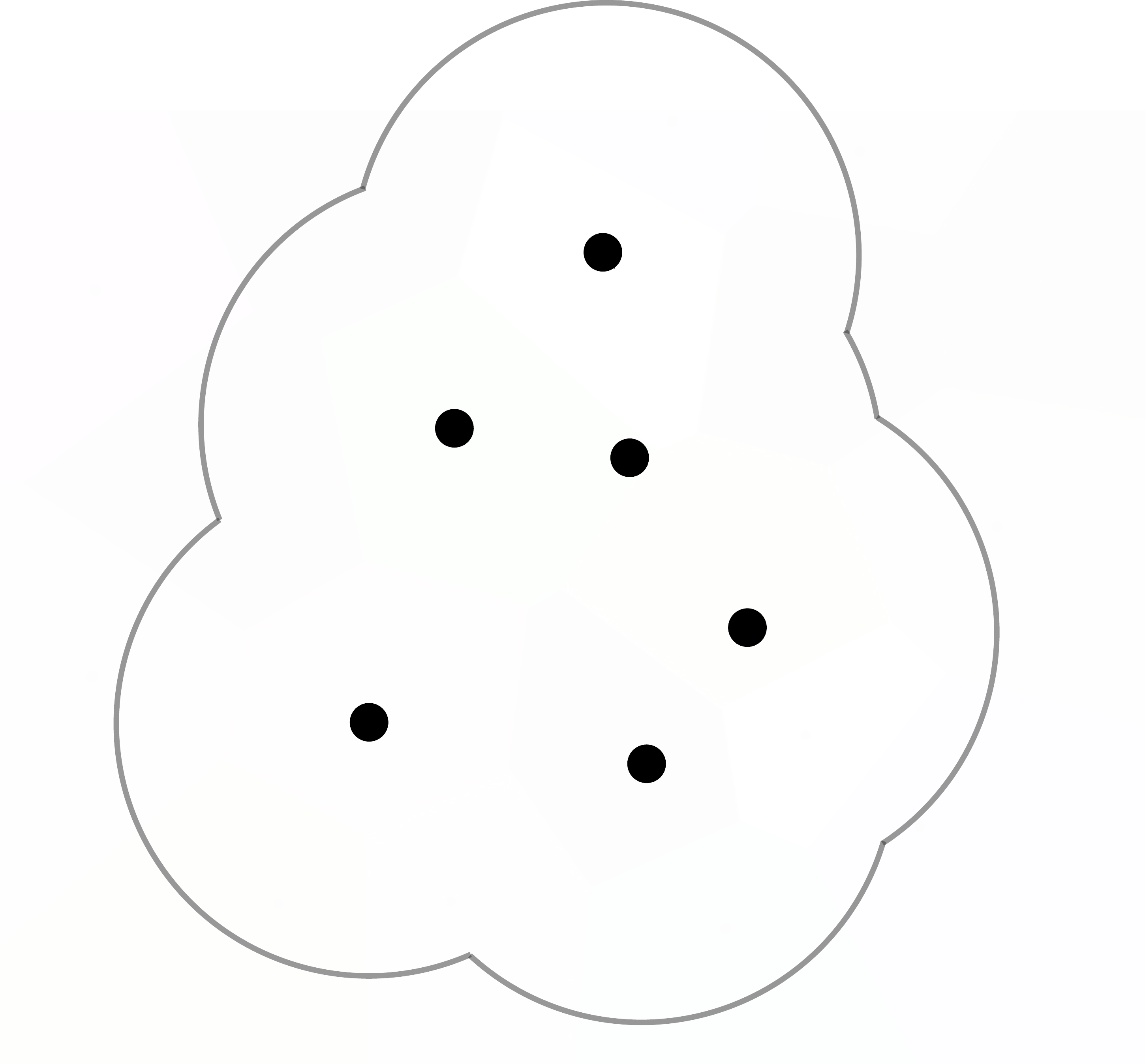}&
\includegraphics[width=0.275\textwidth]{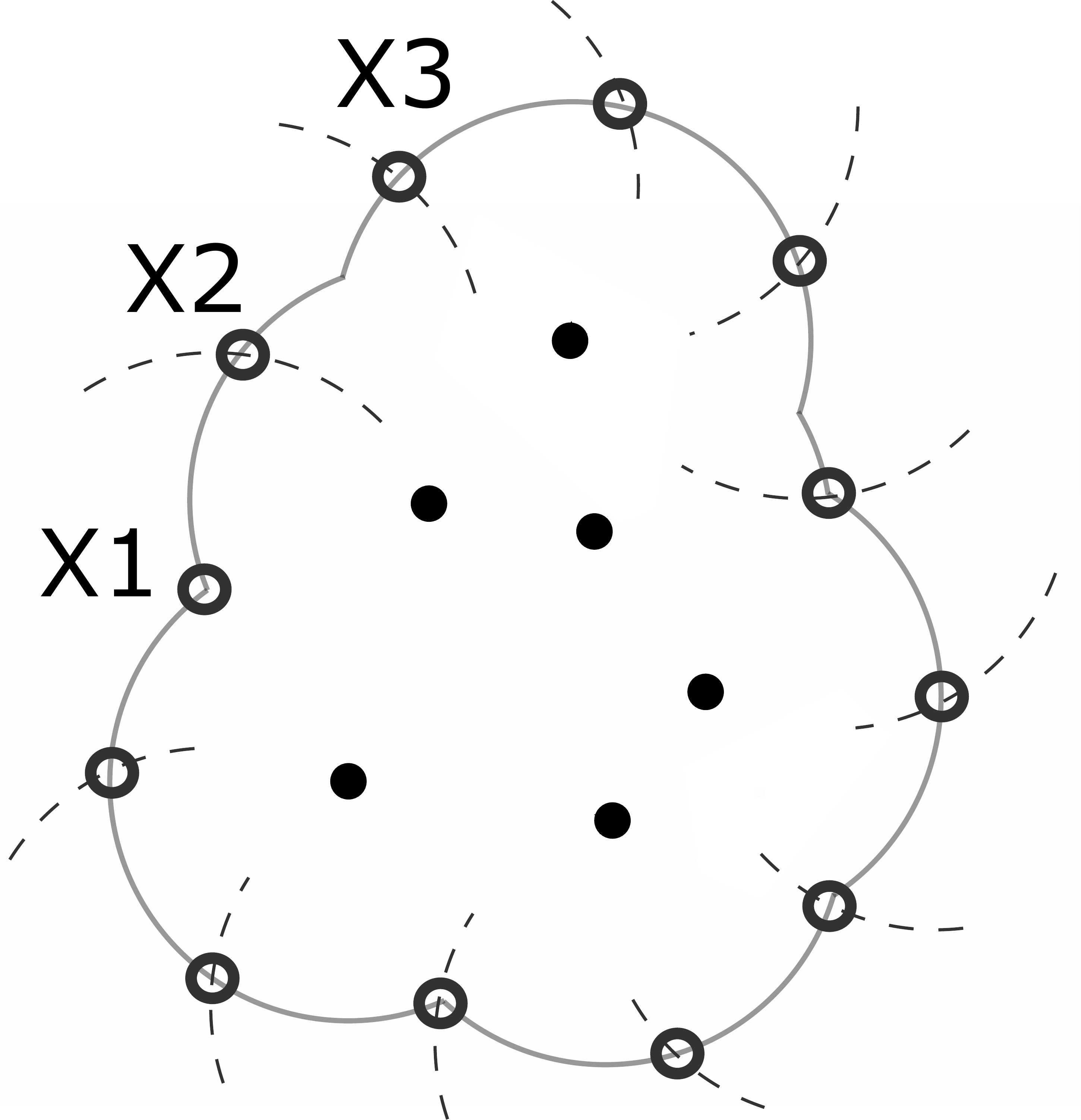}\\
d&e
\end{tabular}
\vspace*{15px}
\end{center}

\noindent{\bf Figure 3.} One iteration of an infinite net construction: a) the given set of AS; b) Delaunay triangulation; c) and d) surrounding equidistant construction; e) placing new layer of virtual nodes of the infinite net on the equidistant
\vspace*{15px}

\newpage

\centering


\begin{tabular}{ | c || c | c | c | c | c || c | c |}
\hline
  species & $\alpha_1$ & \ldots & $\alpha_\Psi$ &  \ldots & $\alpha_\Omega$ & $T(\mathscr{A})$ & $K^0(\mathscr{A})$\\\hline\hline
  \ldots & \ldots & \ldots & \ldots& \ldots & \ldots & \ldots & \ldots\\
\end{tabular}
\vspace*{15px}

\noindent{\bf Figure 4.} General form of the DB

\end{document}